# Probing exciton interaction with a spatially defined charge accumulation layer in the organic semiconductor Diindenoperylene.

N.H. Hansen[1,*], C. Wunderlich[1], A.K. Topczak[1,2] and J. Pflaum[1,3,†]

[1]Julius-Maximillians University, Am Hubland, 97074 Würzburg, Germany

[2]Center for Nanosystems Chemistry, Am Hubland, 97074 Würzburg, Germany

[3]ZAE Bayern, Am Hubland, 97074 Würzburg, Germany

Abstract:

We present an investigation of the microscopic interplay between excitons and charge carriers by means of combined photoluminescence (PL) and charge carrier transport measurements on organic thin film transistors (OTFT). For this purpose, the prototypical organic semiconductor Diindenoperylene (DIP) was utilized as active material. The OTFT accumulation layer provides a spatially defined interaction zone for charges and photo-generated excitons leading to a PL intensity reduction of up to 4.5%. This effect correlates with the accumulated hole carrier density and provides a lower estimate of about $1.3 \cdot 10^{-10}$ cm$^3$/s for the cross-section of non-radiative exciton-hole processes. It is rationalized that these processes are preferentially mediated by trapped holes.

Excitons formed by electrostatically bound electron and hole pairs constitute fundamental excitations in organic semiconducting (OSC) materials. Generated either by photon absorption or by mutual capturing of charge carriers of opposite polarity, excitons play a key role in organic photo-voltaic cells (OPVs) or organic light emitting diodes (OLEDs), respectively.[1,2,3,4] Vice versa, exciton loss mechanisms corrupt the performance of opto-electronic devices and therefore have to be minimized.[5,6] Whereas strong Coulomb-binding energies of up to 1 eV are able to stabilize paired electron-hole states in organic semiconductors against thermally activated dissociation, interaction between excitons or with single charge carriers might lead to singlet exciton losses by various pathways. Amongst which are exciton-exciton annihilation, where by non-radiative recombination of one electron-hole pair the remaining pair gains sufficient energy to overcome the Coulomb-binding energy required for dissociation, or charge carrier exchange accompanied by the formation of an energetically lower-lying triplet state.[7,8,9] Despite their importance, studies on the exciton loss mechanisms in neat organic single layers together with an estimation of the pivotal cross-sections are rather the exception, due to difficulties in precisely controlling the underlying boundary-conditions and in achieving a response of suited size to quantify the occurring loss processes.

In this paper we focus on microscopic exciton quenching mechanisms induced by charge carriers and their investigation by photoluminescence (PL) quenching measurements on organic thin film transistor (OTFT) devices. This novel approach offers two essential advantages compared to studies on real opto-electronic devices. At first, material inherent mechanisms can be identified without superimposed effects at non-ideal heterojunction interfaces, such as locally inhomogeneous electric field distributions. Moreover, the choice of contact metal, in principle, enables discrimination of recombination effects induced by electrons from those related to holes. At second, the tunable charge density in the TFT accumulation layer provides an additional degree of freedom to systematically influence charge carrier mediated exciton annihilation processes. Complementary studies of the PL quenching at different operational regimes of the transistor might unveil the origin of the participating charges, in particular if they are free or spatially immobilized, e.g. by traps.

As organic semiconducting material the polyaromatic hydrocarbon Diindenoperylene (DIP) was chosen. Besides its exceptional exciton diffusion properties in correlation with a long-range ordered thin film structure, which are necessary for the excitons to reach the accumulated charges, this compound has already proven successfully in OTFT applications as well as in photovoltaic cells.[10,11,12,13] Furthermore, the low optical absorption of DIP proves beneficial for the intended studies on exciton-charge carrier interaction since it avoids pronounced quenching contributions by bimolecular recombination.

Thin film transistors were prepared via a five-step shadow mask process in bottom-gate top-contact geometry as described by Klauk et al.[14] Illustrated in Fig. 1 a), a 60 nm thick aluminium film was evaporated under HV (base pressure $10^{-6}$ mbar) at a rate of 1.8 nm/s on top of a Si wafer covered by 200 nm thick thermally grown $SiO_2$ (roughness below 0.4 nm). The native AlOx. surface layer was further increased by oxygen plasma treatment yielding a total thickness of 10.8 nm as confirmed by X-ray reflectivity (XRR) measurements. Thereafter, a self-assembled monolayer of pentadecylfluoro-octadecylphosphonic acid (FOPA) was grown from 2-propanol solution on-top of the aluminum oxide overnight. The pronounced OH-termination of the AlOx. surface upon oxygen plasma treatment promotes the binding of the phosphonic acid groups of FOPA, therefore facilitating the formation of a high quality SAM with respect to film homogeneity and packing density. The effective FOPA film thickness of 2.2 nm has been verified by XRR measurements. By its chain length of 18 carbon atoms, the FOPA SAM enables reliable transistor operation in a voltage regime of up to 3 V and, even more important, its band-gap of about 9 eV prevents injection of parasitic charges by the applied gate-field or by the photo-effect upon optical illumination of the OTFT (see below). Subsequent to gate preparation, DIP layers with thicknesses of up to 75 nm were grown by thermal evaporation under HV at a rate of 0.025 nm/s. The upper thickness limit was chosen according to X-ray diffraction studies yielding a DIP crystallite extension along the surface normal, i.e. along the c'-direction, of similar size and therefore ensuring unhindered excition transport along the vertical direction. Furthermore, assuming an index of refraction along the surface normal of $n_\perp = 1.7$ optical excitation at $\lambda_{exc.} = 532$ nm corresponds to an effective wavelength of $\lambda_{eff.} = 300$ nm within the molecular layer. Thus, all DIP thicknesses comply with the condition $d_{DIP} < \lambda/4$ and therefore, interference effects can be disregarded. Finally, the source- and drain-top-contacts were made by 70 nm Au thermally deposited in HV at a fast rate of 2 nm/s to avoid metal penetration into the organic semiconductor layer underneath.[15,16]

To measure the voltage dependent photoluminescence and thereby, to estimate the amount of excitons quenched under charge accumulation, the DIP channel of the OTFTs was optically excited by a frequency doubled cw-Nd:YAG laser ($\lambda_{exc.}$ = 532 nm) at a nominal power of 0,2 mW. The laser intensity remained constant during transistor operation and was kept sufficiently low to guarantee a linear correlation with the generated exciton density, i.e. to avoid bimolecular contributions to the PL signal. This is corroborated by the fact that, assuming an exponential generation profile according to Lambert-Beer's law, at full power the 2D projected area density of excitons amounts to at most $1/\mu m^2$, i.e. one exciton per $10^6$ DIP molecules, which yields to an average exciton spacing substantially larger than the distance covered by the reported exciton diffusion length of about 100 nm along c'-direction.[10] The PL emitted by the DIP transport layer was detected by a CCD camera (QE > 90% in the measured spectral range) in combination with a spectrometer. To avoid degradation of the organic layer upon photo-excitation all measurements have been performed under nitrogen atmosphere. At first, a PL measurement on the unbiased transistor was carried-out as reference for the subsequent studies at various $V_G$, i.e. various charge densities of the accumulated layer at the DIP/FOPA interface. To evaluate the PL emission at different regimes of TFT operation the relative quenching Q, defined by the difference between PL at $V_G$ = 0 ($PL^{nQ}$) and PL at $V_G \neq 0$ ($PL^Q$) normalized to the former has been employed,

$$Q = 1 - \frac{PL^Q}{PL^{nQ}} \quad . \qquad (1)$$

Thus the relative quenching can be determined as a function of the gate voltage and consequently of the accumulated charge density using the areal capacitance $c_i$ = 0.44 µF/cm², which has been derived by the film thicknesses of the SAM (2.2 nm) and the AlOx. layer (10.8 nm) and assuming $\varepsilon_{SAM}$ = 2.5 and $\varepsilon_{AlOx}$ = 9.4 for the dielectric constants, respectively.

By determining the relative quenching the measurements become independent of the absolute PL intensity, which could differ for each sample, due to variation of the DIP film thickness. In addition, bimolecular processes as well as quenching at crystallographic defects like grain boundaries or impurities are not accounted for by the relative measurement since they are equally present regardless of the applied gate voltage.

The laser spot of 600 µm diameter was focused on the transistor channel (L = 100-200 µm, W = 1.5 mm). Illuminated parts outside the accumulation zone contributed a voltage-independent background intensity of about 30% to the total PL signal at $V_G$ = 0. We therefore consider the data deduced by PL quenching as a lower estimate. Prior to PL detection the equilibrium of charge distribution inside the DIP semiconducting channel was assured by sufficiently long delay and integration times during ramping of the applied voltage.

To evaluate parasitic charge generation by photo-release from the FOPA/AlOx. gate dielectric we compared the gate currents of transfer curves measured in the dark and under illumination, indicating no noticeable photo-generated charge contribution to the gate current. Furthermore, possible effects by the concentration gradient and the diffusion of excitons in the organic transport layer, the PL quenching were investigated on OTFTs with different DIP layer thicknesses between 25 and 75 nm. Yet no distinct influence of the exciton diffusion could be estimated, indicating this thickness regime to be smaller than the exciton diffusion length reported for DIP.[10]

The drain current increases upon illumination by 10% with respect to the current of 2 µA created by the field accumulated charges. Furthermore, turn-on and threshold voltages, determined from the saturation regime, are shifted towards more positive values, as can be seen in Fig. 1b. To appraise possible effects by field induced exciton dissociation in the DIP channel it has to be considered that any out-of-plane component of the electric field is compensated by the charge accumulation layer. Vice versa, the in-plane field between source and drain of $E_{max}$ = 2·10$^2$ V/cm is too small by orders of magnitude to compete with the critical field for exciton dissociation of $E_c$ = 5·10$^6$ V/cm which has been derived using

$$E_c = \frac{E_B}{er} \quad , \qquad (2)$$

with a DIP exciton binding energy of $E_B$ = 0.5 eV[17] and a dipole extension of r = 1 nm; e is the elementary charge. Therefore, field-assisted exciton dissociation can be discarded. Instead, we attribute the observed shifts to an increased hole injection into the organic semiconductor upon illumination leading to an effective decrease in the trap density as discussed by Orgiu et. al.[18]

The PL intensity and its variation with gate voltage are displayed in the inset of Fig. 2 for a 60 nm thick DIP transistor channel. Accordingly, the relative quenching can be extracted at different drain voltages as a function of the applied gate voltage as shown in Fig. 2. An almost linear increase of the relative quenching with gate-voltage is apparent for small values of $|V_G|$, followed by saturation for $|V_G| \geq 1.5$ V. In this regime, the maximum quenching reaches up to 4.5% and, therefore, provides an upper limit for non-radiative exciton-hole interaction with respect to the applied gate voltage.

As the quenching increases linearly with gate voltage above the threshold voltage, we can conclude on the nature of charge carriers interacting with the photo-generated excitons. Up to the threshold voltage, the injected charges, in this case holes, behave like quasi-free, i.e. their transport is significantly affected by trap states near the band edge. Due to the continuous filling of trap states upon raising of the Fermi level, traps behave energetically more shallow and therefore trapping becomes less efficient.[19] As the threshold voltage marks the cross-over region where most of the traps are filled and the injected holes behave as free charge carriers, we can relate the observed PL quenching mainly to quasi-free, i.e. trapped charge carriers. To corroborate this model we compare the saturation voltage $V_G(Q_{sat})$ with the threshold voltage $V_{th,sat}$ under illumination at various drain voltages (Figure 3). Both voltages increase linearly by a slope of 0.2 yet $V_G(Q_{sat})$ is shifted towards more negative values by -0.8 V. Therefore, quenching is not only mediated by occupied deep traps but also by occupied shallow traps that promote hole transport in this voltage regime. As soon as the saturation voltage is reached, holes are trapped on time scales being too short for contributing to the quenching. Hence, the increasing free charge carrier density does not significantly affect the quenching efficiency.[20] Moreover, the similar slope of $V_G(Q_{sat})$ and $V_{th,sat}$ as a function of drain voltage confirms our assumption of trapped holes being the origin of the quenching. An increasing drain voltage compensates the gate potential near the drain contact further (pinch-off) and therefore effectively decreases the occupied trap density, which in turn is directly linked to $V_G(Q_{sat})$ and $V_{th,sat}$. Our observation coincides qualitatively with previous fluorescence quenching studies on anthracene single crystals[20] indicating that exciton interaction with trapped holes constitutes the major non-radiative loss channel, whereas PL decay remains constant upon increase of the free charge carrier density.

To summarize the microscopic decay channels in more detail Fig. 4a depicts the relevant exciton-related processes: First, an exciton is photogenerated upon light absorption within the DIP layer (step 1.). Under steady-state conditions an exponential distribution of the exciton density will be established inside the organic layer. Secondly, within the concentration gradient the exciton will diffuse to the charge accumulation layer (step 2.) or radiatively recombine within its lifetime (step 3.). However, if the exciton is able to reach the accumulation layer at the DIP/FOPA interface it may recombine radiatively as well as non-radiatively (step 4.). Possible recombination processes contributing to step 4. are listed in detail in Fig. 4b. As we do not observe additional PL quenching contributions when the density of free charge carriers is enhanced further, all quenching mechanisms based on the interaction between excitons and free holes can be disregarded (right branch of Fig. 4b). Therefore, trapped positive carriers can be identified to mediate the non-radiative quenching of the photogenerated excitons (left branch in Fig. 4b).

Estimating a total of $10^{-5}$ excitons per charge carrier for the applied laser intensity and with respect to the geometrical dimensions of the OTFT channel, we determined a lower limit for the effective cross-section of exciton-hole interaction

$$K = \frac{\Delta Q}{\Delta n_{trap} \tau_{ex}} = \frac{\Delta Q}{\Delta V_G c_i \tau_{ex}} \quad , \quad (3)$$

With ΔQ/ΔV$_G$ describing the change in relative quenching upon variation of the gate voltage, while $\tau_{ex}$ is the exciton lifetime. Heilig et al.[21] have estimated an exciton lifetime of several ns for DIP at cryogenic temperatures. To account for our room-temperature measurements and the fact that most excitons spend only a fraction of their lifetime inside the accumulation zone we assumed $\tau_{ex}$ = 1 ns for further calculations. We deduced the lower limit of exciton – hole interaction to be about 1.3·10$^{-10}$ cm$^3$/s. Though no reliable data on this quantity are available for thin film systems so far, the comparison with single crystal PL quenching data, e.g. 8·10$^{-10}$ cm$^3$/s for naphthalene,[22] renders a sufficiently good agreement with our data. Therefore, estimation of this property by voltage dependent luminescence quenching in thin film transistor devices provides a method applicable for a variety of luminescent compounds and for technologically relevant film thicknesses.

In summary we presented a novel approach for the investigation of exciton-charge carrier interaction in organic semiconductors. By utilizing TFT structures to create a well-defined charge carrier accumulation zone at the semiconductor/insulator interface we measured the photoluminescence quenching as a function of the accumulated charge carrier density. In the course of these studies, two different quenching regimes could be identified where from the linear regime the quenching could be assigned to localized holes in the accumulation zone. From the maximum quenching efficiency of 4.5% we deduced an lower limit of the effective cross section for exciton – localized hole interaction of about 1.3·10$^{-10}$ cm$^3$/s which matches that reported for low-weight molecular single crystals.

Acknowledgements:

We gratefully acknowledge the generous financial support by the Bavarian State Ministry of Science, Research, and the Arts within the Collaborative Research Network "Solar Technologies go Hybrid". This work was supported by the BMBF-Project GREKOS and the DFG-Project SPP1355.

[*]nhansen@physik.uni-wuerzburg.de

[†]jpflaum@physik.uni-wuerzburg.de

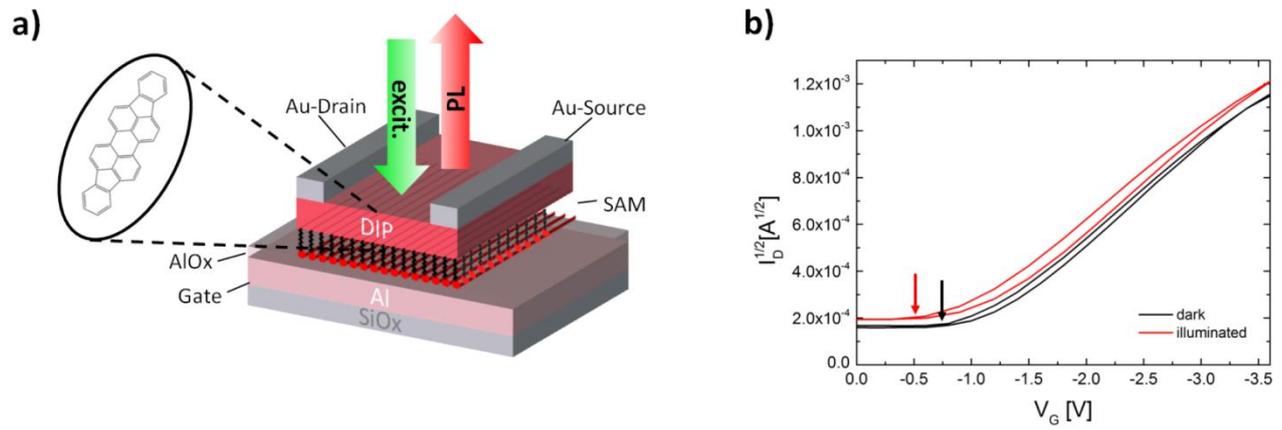

FIG. 1: a) DIP thin film transistor geometry, utilized for photoluminescence quenching studies. b) OTFT transfer characteristics measured in the dark and under 532 nm illumination. As indicated by the arrows a shift towards smaller threshold voltages occurs upon illumination whereas the absolute drain current remains almost constant (uncertainty of the data below 5%).

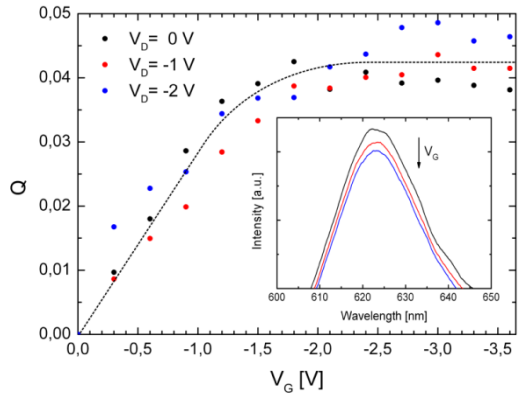

FIG. 2: Relative photoluminescence quenching, $Q$, of a 60 nm thick DIP transistor as a function of gate-voltage, $V_G$, for three different drain voltages, $V_D$. The independence of $Q$ on $V_D$, i.e. on the density of free charges within the OTFT channel and its monotonous rise with increasing gate voltage hint at the exciton interaction at the hole accumulation layer. The dashed line presents a guide-to-the-eye to the data. The inset illustrates the effect of $V_G$ variation on the PL intensity of the DIP S(1;0) transition. The decrease in peak intensity corresponds to an overall quenching of about 4.5 %.

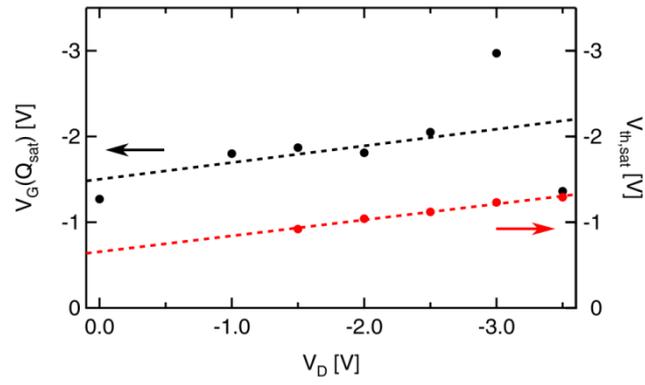

FIG. 3: Threshold voltage $V_{th,sat}$ and transition voltage $V_G(Q_{sat})$ under illumination as functions of the applied drain voltage. The similar slope of both quantities (dashed lines) evidences the same origin, namely the density of occupied traps.

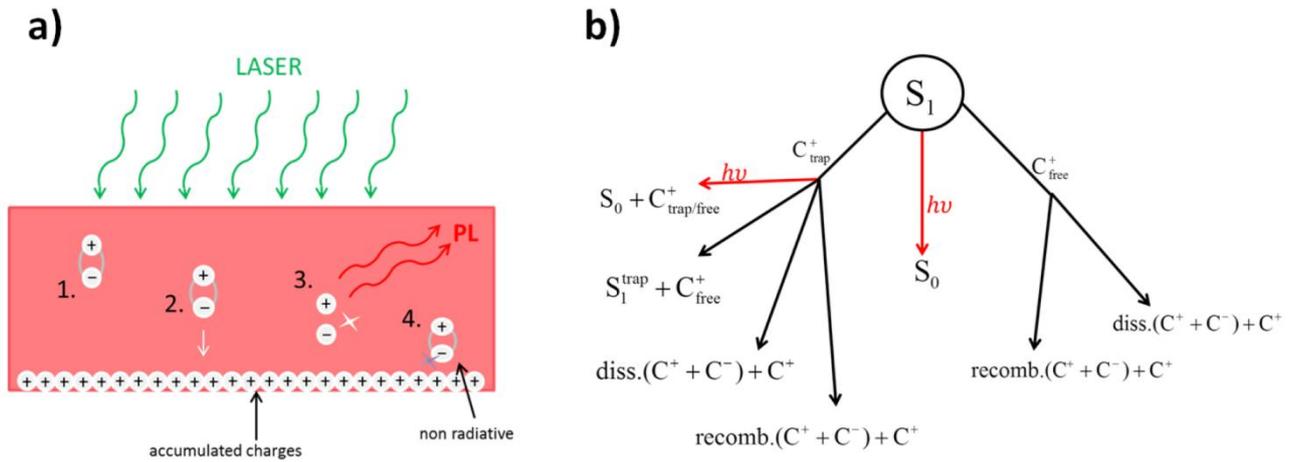

FIG. 4: a) Illustration of the relevant excitonic processes within the OTFT configuration. After photo-generation (1.), exciton diffusion is steered by the concentration gradient towards the hole accumulation layer (2.). Upon diffusion, either radiative recombination (3.) or non-radiative processes (4.), induced at the charge accumulation layer and yielding PL quenching, can occur. b) Possible recombination scenarios leading to either PL emission (red arrows) or quenching (black arrows).